\UseRawInputEncoding

\documentclass[aps,amsmath,amssymb,floatfix,reprint,citeautoscript,
noeprint,superscriptaddress,twocolumn]{revtex4-2}
\usepackage{dsfont}
\usepackage{lipsum} 
\usepackage{bibentry}
\usepackage[english]{babel}
\usepackage[dvipsnames]{xcolor}
\usepackage{graphicx}
\usepackage[caption=false]{subfig} 
\usepackage{amsmath,amssymb,bm}
\usepackage[version=3]{mhchem}
\usepackage{verbatim}
\usepackage{multirow}
\usepackage{dcolumn}
\usepackage{float}
\usepackage{nicefrac}
\usepackage{siunitx}
\usepackage{booktabs}
\usepackage{chemformula}
\usepackage{wrapfig}
\usepackage{enumitem}  
\usepackage{transparent}
\usepackage[colorlinks,allcolors=black,citecolor=blue,urlcolor=blue]{hyperref}
\usepackage{tcolorbox}
\usepackage{relsize}
\hypersetup{
    citecolor=Turquoise,
    colorlinks=true,
    linkcolor=black,    
    urlcolor=Turquoise,
    }

\newcommand{\mbf}[1]{\boldsymbol{\mathit{#1}}}
\newcommand{\mrm}[1]{\mathrm{#1}}
\newcommand{\mcl}[1]{\mathcal{#1}}

\newcommand{\tcr}[1]{\textcolor{magenta}{#1}}

\newcommand{\etal}{\emph{et al.}}

\newcommand{\erf}{\mathrm{erf}}

\usepackage{sansmathfonts}
\usepackage[justification=centerlast, font={sf}, labelfont={bf}]{caption}
\begin{document}

\title{A first principles approach to electromechanics in liquids}

\author{Anna T. Bui}
\affiliation{Yusuf Hamied Department of Chemistry, University of
  Cambridge, Lensfield Road, Cambridge, CB2 1EW, United Kingdom}
\affiliation{Department of Chemistry, Durham University, South Road,
  Durham, DH1 3LE, United Kingdom}

\author{Stephen J. Cox}
\email{stephen.j.cox@durham.ac.uk}
\affiliation{Department of Chemistry, Durham University, South Road,
  Durham, DH1 3LE, United Kingdom}

\date{\today}


\begin{abstract}
Electromechanics in fluids describes the response of the number
density to electric fields, and thus provides a powerful means by
which to control the behavior of liquids. While continuum approaches
have proven successful in describing electromechanical phenomena in
macroscopic bodies, their use is questionable when relevant length
scales become comparable to a system's natural correlation lengths, as
commonly occurs in, e.g., biological systems, nanopores, and
microfluidics. Here, we present a first principles theory for
electromechanical phenomena in fluids. Our approach is based on the
recently proposed hyperdensity functional theory [Samm\"{u}ller \etal,
Phys. Rev. Lett. \textbf{133}, 098201 (2024)] in which we treat the
charge density as an observable of the system, with the intrinsic
Helmholtz free energy functional dependent upon both density and
electrostatic potential. Expressions for the coupling between number
and charge densities emerge naturally in this formalism, avoiding the
need to construct density-dependent and spatially-varying material
parameters such as the dielectric constant. Furthermore, we make our
theory practical by deriving a connection between hyperdensity
functional theory and local molecular field theory, which facilitates
machine learning explicit representations for the free energy
functionals of systems with short-ranged electrostatic interactions,
with long-ranged effects accounted for in a well-controlled mean field
fashion.
\end{abstract}

\maketitle

\section{Introduction}

In the presence of an electric field, all fluids will undergo some
degree of polarization---electron clouds will distort, permanent
electric dipoles will reorient, and mobile charges will migrate.
Electromechanics describes the less obvious phenomenon in which the
local number density $\rho(\mbf{r})$ may also respond, arising from a
coupling between $\rho(\mbf{r})$ and the charge density $n(\mbf{r})$
of the fluid.  This effect is particularly relevant in the
presence of strong, spatially-varying electric fields, which are
especially prevalent near interfaces, e.g., at liquid--liquid
interfaces in emulsion droplets \cite{Eow2001}, liquid--solid
interfaces in supercapacitors \cite{Salanne2016} and porous biological
membranes
\cite{Lee1994}. Electromechanical coupling also plays an important role in
capillary phenomena of polar fluids \cite{Zhang2020pccp, bui2025efg}.
Not only of fundamental interest, electromechanics also provides a set
of powerful tools for controlling the behavior of
liquids. Experimental studies have demonstrated how electric fields
can drive phase transitions \cite{Tsori2004}, modify wetting
properties \cite{Hayes2003, McHale2011} and influence fluid
transport \cite{Jones2001}.  Faithfully capturing such
electromechanical coupling is, therefore, essential for any general
theoretical framework that aims to describe the structure and
thermodynamics of inhomogeneous liquids.

Generally speaking, our theoretical understanding of electromechanical
effects is rooted in continuum theories, in which material properties
are encoded in constitutive relations that lead to expressions for the
pressure tensor and force balance equations.
For example, using work
arguments, Landau and Lifshitz \cite{LandauLifshitzBook} derived a 
contribution to the force
density in a fluid dielectric medium due to electrostriction,
\[\frac{1}{8\pi}\nabla\left(\rho(\mbf{r})\bigg(\frac{\partial\epsilon}{\partial\rho}\bigg)\frac{|\mbf{E}(\mbf{r})|^2}{2}\right) - \frac{|\mbf{E}(\mbf{r})|^2}{8\pi}\nabla\epsilon(\mbf{r}),\] where $\epsilon$ is the
density-dependent dielectric constant and $\mbf{E}$ is the Maxwell
electric field.  While such continuum-based approaches have proven
successful for describing macroscopic systems, complications arise
when relevant length scales in the system become comparable to natural
correlation lengths.  Such scenarios occur frequently in confined
geometries and impact physical behavior even in the absence of
electric fields; a well-known example is capillary condensation in
which a phase that is only metastable for the bulk fluid is
stabilized by the presence of confining boundaries separated by a
finite distance \cite{Evans1990}.  In the case of electromechanical
coupling, additional sources of competing length scales can arise from
spatially varying electric fields, which may be as a result of careful
design by the experimentalist \cite{Freedman2016,Hayler2024, McHale2011},
or due to natural
heterogeneity and defects such as in porous carbon
electrodes \cite{Liu2024}.  Such obvious complexity strongly motivates
the need for a theoretical framework with a direct connection to the
underlying microscopic Hamiltonian, and from which electromechanical
coupling is an emergent phenomenon rather than a mere
postulation. Ideally, any theoretical framework should also lend
itself to practical computation.

In this article we present such a theoretical framework. Our approach
is rooted in hyperdensity functional theory
(hyper-DFT) \cite{Sammuller2024hdft}, in which we treat the charge
density as an observable of the system, with a functional dependence
on both the number density field and external electrostatic potential.
In this approach, expressions for susceptibilities and direct
correlation functions relevant to electromechanics emerge naturally,
and can be expressed in terms of fluctuations of the system either
directly as covariances with the one-body density operator, or via
hyperdensity Ornstein-Zernike relations. Important for the practical
implementation of the theory, we also derive the relationship between
hyper-DFT and local molecular field theory (LMFT) \cite{Weeks1998},
generalizing the previous connection between LMFT and classical
density functional theory (cDFT) made by Archer and
Evans \cite{Archer2013}. This relationship allows us to reformulate
our expressions with reference to a system with short-ranged
electrostatic interactions and an effective external electrostatic
potential, which opens the door to applying machine learning
techniques
that depend upon the locality of direct
correlations \cite{Sammuller2023}.

The rest of the article is organized as follows. In
Sec.~\ref{sec:hyperDFT}, we describe the formulation of the theory,
formalizing the treatment of the charge density as a hyperdensity
functional.  In Sec.~\ref{sec:electromechanics}, we discuss the
electromechanical coupling that emerges from the theory,
including functional relationships for the density and charge
responses to both non-electrostatic and electrostatic external
potentials. In Sec.~\ref{sec:LMFT}, we establish the connection
between hyper-DFT and LMFT for fluids with electrostatic
interactions. We give our conclusions and outlook in
Sec.~\ref{sec:conclusions}.

\section{Treating charge density as an observable with hyper-DFT} \label{sec:hyperDFT}

As set out above, we seek to maintain a direct connection to the
microscopic Hamiltonian,
\begin{equation}
  \label{eqn:H}
\hat{\mcl{H}}(\mbf{P}^N,\mbf{R}^N) = \hat{\mcl{K}}(\mbf{P}^N) + \hat{\mcl{U}}(\mbf{R}^N) + \hat{\mcl{V}}_{\rm ext}(\mbf{R}^N),
\end{equation}
for a fluid of $N$ particles of mass $M$. A particle itself may comprise $s$
interaction sites (not necessarily charged), and the notation
$\mbf{P}^N$ should, therefore, be understood to indicate both the
center-of-mass momentum of the particles, and the momenta associated
with internal degrees of freedom. Similarly, $\mbf{R}^N$ indicates
both the particle centers, which for particle $i$ we indicate by
$\mbf{r}_i$, and the internal configuration of the particle. The
instantaneous values of the kinetic and interparticle potential
energies are denoted by $\hat{\mcl{K}}$ and $\hat{\mcl{U}}$,
respectively, and the ``mechanical'' external potential takes a
one-body form, $\hat{\mcl{V}}_{\rm ext}(\mbf{R}^N) = \sum^N_i V_{\rm
ext}(\mbf{r}_i)$, 
and is purely
non-electrostatic. We will work in the grand canonical ensemble, such
that the relevant partition function is
\begin{equation}
\label{eqn:Xi0}
\Xi_0=\mrm{Tr}\,\exp\left[ - \beta \left(\hat{\mcl{H}} - \mu N\right)\right],
\end{equation}
where $\beta = 1/k_{\rm B}T$, with $k_{\rm B}$
denoting the Boltzmann constant and $T$ the temperature, 
and $\mrm{Tr}$ indicates the classical ``trace'' operator
\begin{equation}
\label{eqn:Tr}
\mrm{Tr} = \frac{1}{N!h^\alpha}\sum_{N=0}^\infty\int\!\mrm{d}\mbf{P}^N\!\int\!\mrm{d}\mbf{R}^N,
\end{equation}
with $h$ denoting the Planck constant, and where $\alpha$ is the total
number of degrees of freedom of the system. ($\alpha$ will depend upon
$N$, $s$, and the details of the intramolecular degrees of
freedom, including any constraints that are imposed.)
An important conceptual point for what follows is that, having
specified the grand canonical ensemble, it is clear that a particle is
defined as a body for which the chemical potential $\mu$ is controlled
in the reservoir.

To describe interparticle electrostatic interactions, we introduce the
instantaneous charge density operator $\hat{n}(\mbf{r},\mbf{R}^N)$,
which, for a given configuration of the system, returns the charge
density at position $\mbf{r}$. 
We have deliberately left its form
unspecified; while in certain cases $\hat{n}$ depends ``trivially'' on $\mbf{R}^N$
(e.g., simple point charge models), in general the relationship can be
complicated. 
The potential energy of the system can be written as
\begin{equation}
    \hat{\mcl{U}}(\mbf{R}^N) =  \hat{\mcl{U}}_{\mrm{ne}}(\mbf{R}^N)
    +  \frac{1}{2} \int\!\!\mrm{d}\mbf{r}\!\int\!\!\mrm{d}\mbf{r}^\prime\, \frac{\hat{n}(\mbf{r},\mbf{R}^N)\hat{n}(\mbf{r}^\prime, \mbf{R}^N)}{|\mbf{r}-\mbf{r}^\prime|},
\end{equation}
which also defines the non-electrostatic contribution to the potential
energy $\hat{\mcl{U}}_{\rm ne}$.

We now consider the action of an external electrostatic potential
$\phi(\mbf{r})$. In this case, the partition function is modified in a
straightforward manner,
\begin{equation}
\label{eqn:Xi_phi}
\Xi_{\phi}=\mrm{Tr}\,\exp\left[ - \beta \left(\hat{\mcl{H}} + \!\int\!\!\mrm{d}\mbf{r}\,\phi(\mbf{r})\hat{n}(\mbf{r}) -\mu N\right)\right].
\end{equation}
The grand potential takes the usual form,
\begin{equation}
\Omega_\phi = -k_{\rm B}T\ln\Xi_{\phi},
\end{equation}
from which the equilibrium average charge density follows from
functional differentiation,
\begin{equation}
\label{eqn:AvgChg-1}
    n(\mbf{r}) \equiv \langle\hat{n}(\mbf{r}, \mbf{R}^N)\rangle_\phi = \frac{\delta\beta\Omega_{\phi}}{\delta\beta\phi(\mbf{r})},
\end{equation}
where $\langle \cdots \rangle_{\phi}$ denotes an ensemble average
corresponding to the partition function defined in
Eq.~$\ref{eqn:Xi_phi}$.  Implicit in the above functional derivative
is that $\mu$, $T$, and $V_{\rm ext}$ are fixed.

The premise of hyper-DFT rests on the observation that $\Xi_\phi$ is
agnostic to whether the system is considered as one with $\phi$
coupling to $\hat{n}$ as an external electrostatic potential, or as
one with a modified interparticle potential energy
\begin{equation}
    \hat{\mcl{U}}_{n}(\mbf{R}^N) =  \hat{\mcl{U}}(\mbf{R}^N) + \int\!\!\mrm{d}\mbf{r} \, \phi(\mbf{r})\hat{n}(\mbf{r},\mbf{R}^N).
\end{equation}
As the intrinsic Helmholtz free energy has an implicit functional
dependence on the interparticle potential energy, the seemingly
trivial statement above has profound implications when adopting a
density functional approach (see, e.g.,
Ref.~\cite{kampa2024metadft}). In the case at hand, the excess
intrinsic Helmholtz free energy functional $\mcl{F}^{\rm (ex)}_{\rm
intr}[\varrho,\beta\varphi]$ is a unique functional of the number
density $\varrho$, and acquires a functional dependence on the
electrostatic potential $\beta\varphi$, which we now make
explicit. The number density, $\varrho$, is an ensemble average of the
number density operator,
\begin{equation}
\hat{\rho}(\mbf{r},\{\mbf{r}_i\}) = \sum_{i=1}^{N} \delta(\mbf{r}-\mbf{r}_i),
\end{equation}
though the average may not necessarily be that corresponding to the
equilibrium distribution of microstates. For clarity, throughout this
article, we will use $\varrho(\mbf{r})$ and $\varphi(\mbf{r})$ when
denoting functional dependency on any density field and electrostatic
potential, and reserve the notation $\phi(\mbf{r})$ for a
particular external electrostatic potential, and
$\rho(\mbf{r})=\langle\hat{\rho}(\mbf{r},\{\mbf{r}_i\})\rangle_{\phi}$
for the corresponding equilibrium number density. Note that
$\hat{\rho}$ has an explicit dependence on the particle centers,
$\{\mbf{r}_i\}$, but not the internal degrees of freedom of the
particles.

Following the generalization of the Mermin--Evans minimization
principle
\cite{Mermin1965, Evans1979}
presented by Samm\"{u}ller and Schmidt \cite{Sammuller2024review}, we
can introduce the grand potential density functional
$\varOmega[\varrho,\beta\varphi]$, which for a particular
electrostatic potential $\phi$, is minimized by the equilibrium
one-body density,
\begin{equation}
     \frac{\delta\varOmega[\varrho,\beta\phi]}{\delta\varrho(\mbf{r})}\bigg|_{\varrho = \rho} = 0,
\end{equation}
and is equal to the grand potential when evaluated at equilibrium,
\begin{equation}
\Omega_\phi = \varOmega[\rho,\beta\phi].
\end{equation}
From Eq.~\ref{eqn:AvgChg-1}, it immediately follows that $n(\mbf{r})$
can be obtained directly from $\varOmega[\varrho,\beta\varphi]$,
\begin{equation}
n(\mbf{r}) = \frac{\delta\beta\varOmega[\rho,\beta\varphi]}{\delta\beta\varphi(\mbf{r})}\bigg|_{\varphi = \phi}.
\end{equation}
An important  point to emphasize at this stage is that 
$\varOmega[\varrho,\beta\varphi]$ is not a functional of
the charge density. Rather, we are treating the
charge density as an observable of the system that is conjugate to
$\varphi(\mbf{r})$, and finding its average through functional
differentiation. Clearly, the charge density has a functional dependence on
$\varphi(\mbf{r})$. As shown in Ref.~\cite{Sammuller2024review},
any observable can be also be
written as a hyperdensity functional of the one-body density
$\varrho(\mbf{r})$. In what
follows, we will make both of these functional dependencies
explicit. To be concrete, we introduce the one-body charge density functional,
\begin{equation}
\label{eqn:n-Fex}
n^{(1)}(\mbf{r};[\varrho,\beta\varphi]) = \frac{\delta\beta\mcl{F}^{(\rm ex)}_{\rm intr}[\varrho,\beta\varphi]}{\delta\beta\varphi(\mbf{r})},
\end{equation}
from which the charge density at equilibrium follows
\begin{equation}
n(\mbf{r}) = n^{(1)}(\mbf{r};[\rho,\beta\phi]).
\end{equation}

\section{Electromechanical coupling from hyper-DFT} \label{sec:electromechanics}

\begin{figure*}[t]
  \includegraphics[width=0.76\linewidth]{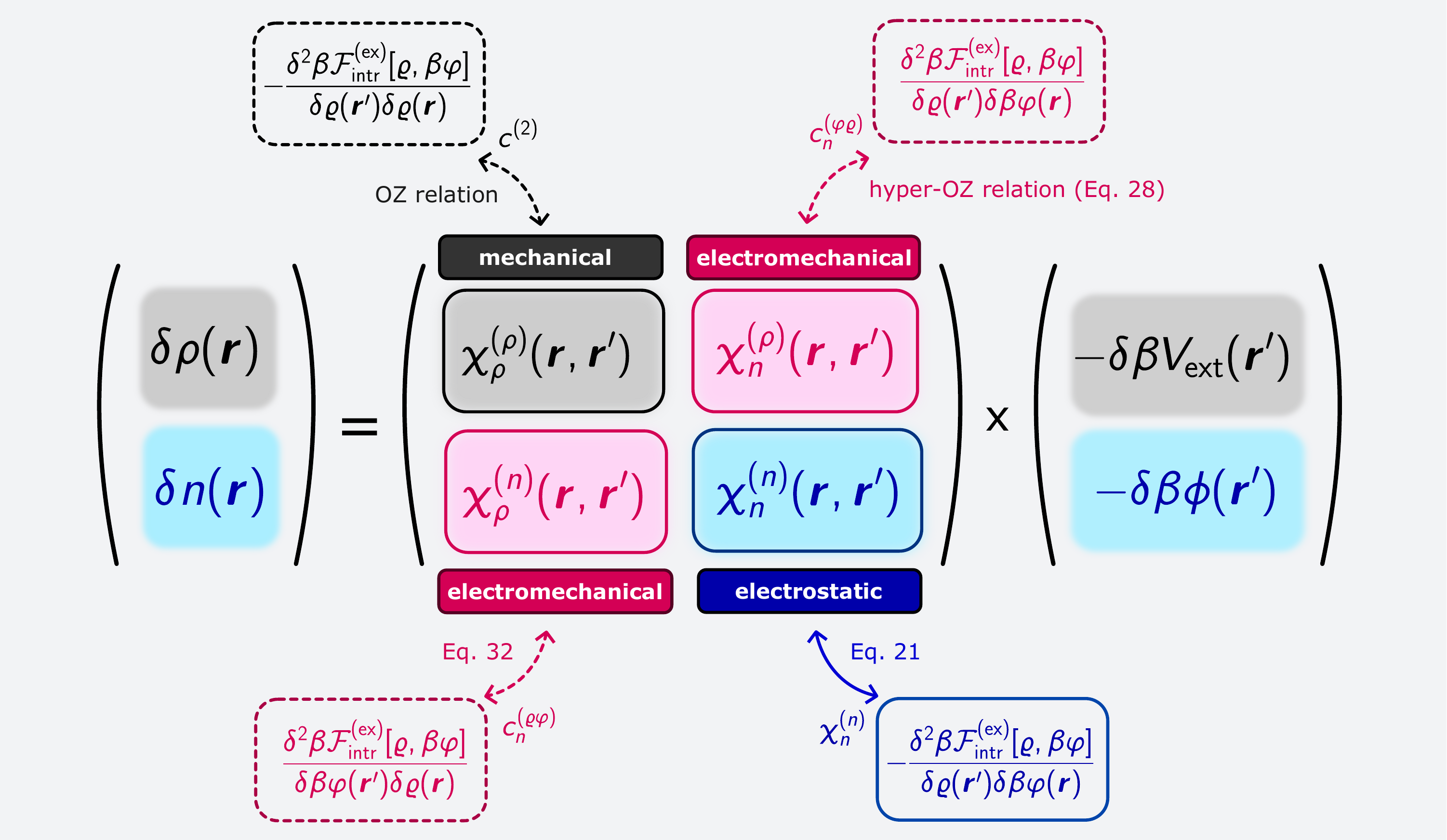} \caption{\textbf{Statistical
  mechanics of response to mechanical and electrostatic perturbations
  for inhomogeneous fluids.} The diagonal elements of the central
  matrix contain the fluid's direct mechanical ($\chi^{(\rho)}_\rho$)
  and electrostatic ($\chi^{(n)}_n$) response functions. Off-diagonal
  elements ($\chi^{(\rho)}_n$ and $\chi^{(n)}_\rho$) describe
  electromechanical responses, and arise from the coupling between
  microscopic number and charge densities. All these response
  functions are connected to the excess free energy hyperdensity
  functional
  $\mcl{F}^{\mrm{(ex)}}_{\mrm{intr}}[\varrho, \beta\varphi]$: in the
  case of $\chi^{(n)}_n$ the relationship is established directly
  through 
  the second functional derivative; in all other cases, the
  relationship is indirect via (hyper)direct correlation
  functionals.}
\label{matrix}
\end{figure*}

Equation~\ref{eqn:n-Fex} provides a firm statistical mechanical
foundation upon which to consider the average charge density. Combined
with the grand potential density functional,
\begin{equation}
\label{eqn:OmegaRhoPhi}
\varOmega[\varrho,\beta\varphi] = \mcl{F}^{\mrm{(id)}}_{\mrm{intr}}[\varrho] + \mcl{F}^{\mrm{(ex)}}_{\mrm{intr}}[\varrho, \beta\varphi] + \int\!\!\mrm{d}\mbf{r}\, \varrho(\mbf{r}) [ V_{\rm ext}(\mbf{r})- \mu],
\end{equation}
where $\mcl{F}_{\rm intr}^{\rm
(id)}[\varrho]=k_{\mrm{B}}T\!\!\int\!\mrm{d}\mbf{r}\,\varrho(\mbf{r})[\ln \zeta^{-1}\Lambda^3\varrho(\mbf{r})-1]$,
is the ideal intrinsic Helmholtz free energy functional, we can
readily derive several insightful statistical mechanical expressions
relevant to electromechanical phenomena. The thermal de Broglie
wavelength is denoted by $\Lambda = h/\sqrt{2\pi M k_{\rm B}T}$,
and $\zeta$ is an intramolecular partition function that depends
upon the details of the particle's internal degrees of freedom. (In the case that each particle
comprises a single site, $\zeta=1$.)


For simplicity, we present the theory for a single
component fluid. In the case of mixtures, an important example of
which is ionic fluids, the theory generalizes in a straightforward
manner, with the excess free energy functional acquiring a dependence
on all density fields \cite{Evans1980, bui2024learningclassicaldensityfunctionals}, i.e.,
$\mcl{F}^{\mrm{(ex)}}_{\mrm{intr}}[\{\varrho_{\gamma}\}, \beta\varphi]$,
where each species $\gamma$ has their own chemical potential
$\mu_{\gamma}$ and non-electrostatic potential $V_{{\rm
ext},\gamma}$.

\subsection{Hierarchy of direct correlation functions}

In the same manner as conventional cDFT, $\mcl{F}^{\rm (ex)}_{\rm
intr}$ acts as a functional generator for a hierarchy of direct
correlation functions pertaining to correlations of the local number
density with itself
\begin{equation}
\label{eqn:DCF-alpha}
c^{(\alpha)}(\mbf{r}_1,\ldots,\mbf{r}_\alpha; [\varrho,\beta\varphi]) =
-\frac{\delta^\alpha \beta\mcl{F}_{\mrm{intr}}^{\rm (ex)}[\varrho,\beta\varphi]}{\delta\varrho(\mbf{r}_{\alpha})\ldots\delta\varrho(\mbf{r}_{1})}.
\end{equation}
Most important for this work are the first two in this hierarchy,
which we will write explicitly for ease of reference. First, the one-body
direct correlation functional is
\begin{equation}
\label{eqn:DCF-1}
    c^{(1)}(\mbf{r}; [\varrho,\beta\varphi]) =
    -\frac{\delta \beta \mcl{F}_{\mrm{intr}}^{\mrm{(ex)}}[\varrho, \beta\varphi]}{\delta \varrho(\mbf{r})}.
\end{equation}
Second, the two-body direct correlation functional is
\begin{equation}
\label{eqn:DCF-2}
    c^{(2)}(\mbf{r}, \mbf{r}^\prime; [\varrho,\beta\varphi]) =
    -\frac{\delta^2 \beta \mcl{F}_{\mrm{intr}}^{\mrm{(ex)}}[\varrho, \beta\varphi]}{\delta \varrho(\mbf{r}^\prime)\delta \varrho(\mbf{r})},
\end{equation}
which can be related to the two-body pair distribution function
exactly through the standard Ornstein--Zernike (OZ)
equation \cite{HansenMcDonaldBook}. The equilibrium density for a
given $\phi$ is obtained by minimizing $\varOmega[\varrho,\beta\phi]$
(see Eq.~\ref{eqn:OmegaRhoPhi}), and satisfies the following
Euler--Lagrange equation,
\begin{equation}
   \rho(\mbf{r}) = \frac{\zeta}{\Lambda^3}\exp\left(-\beta[V_{\rm ext}(\mbf{r}) - \mu] + c^{(1)}(\mbf{r};[\rho,\beta\phi]) \right).
  \label{eqn:Euler-Lagrange-full}
\end{equation}

\subsection{Charge density response to an external electrostatic potential}

When considering the charge density, it is natural to ask how it
responds to a change in external electrostatic potential. We therefore
define the ``charge--charge'' response functional,
\begin{equation}
  \chi^{(n)}_{n}(\mbf{r}, \mbf{r}^\prime; [\varrho,\beta\varphi]) =
  -\frac{\delta n^{(1)}(\mbf{r};  [\varrho, \beta\varphi])}{\delta \beta\varphi(\mbf{r}^\prime)}.
  \label{eqn:chi-nn-def}
\end{equation}
From Eq.~\ref{eqn:n-Fex} we see that this is related to the second
functional derivative of $\mcl{F}^{\rm (ex)}_{\rm
intr}[\varrho,\beta\varphi]$ with respect to $\beta\varphi$,
\begin{equation}
\label{eqn:chi-nn-Fex}
\chi^{(n)}_{n}(\mbf{r}, \mbf{r}^\prime; [\varrho,\beta\varphi]) =
-\frac{\delta^2 \beta \mcl{F}_{\mrm{intr}}^{\mrm{(ex)}}[\varrho, \beta\varphi]}{\delta \beta\varphi(\mbf{r}^\prime)\delta \beta\varphi(\mbf{r})}.
\end{equation}
The microscopic interpretation of $\chi^{(n)}_n$ can be obtained by
considering the partition function (Eq.~\ref{eqn:Xi_phi}),
\begin{equation}
\label{eqn:chi-nn-var}
  \chi^{(n)}_{n}(\mbf{r}, \mbf{r}^\prime; [\rho,\beta\phi]) =
  \big[\langle\hat{n}(\mbf{r})\hat{n}(\mbf{r}^\prime)\rangle_\phi -
       \langle\hat{n}(\mbf{r})\rangle_\phi\langle\hat{n}(\mbf{r}^\prime)\rangle_\phi\big].
\end{equation}
As one might have anticipated, at equilibrium, $\chi^{(n)}_n$ is given by the variance
of the charge density.

\subsection{Number density response to a change in external electrostatic potential}

We also consider how the local number density responds to a change in
external electrostatic potential. We therefore define the 
``number--charge'' response function
\begin{equation}
    \chi^{(\rho)}_{n}(\mbf{r}, \mbf{r}^\prime) =
    \left.-\frac{\delta \rho(\mbf{r}) }{\delta \beta\phi(\mbf{r}^\prime)}   \right\vert_{\beta(V_{\rm ext} - \mu)}.
\end{equation}
Again, we give a microscopic interpretation by considering the
partition function (Eq.~\ref{eqn:Xi_phi}),
\begin{equation}
\chi^{(\rho)}_{n}(\mbf{r}, \mbf{r}^\prime) =
  \big[\langle\hat{\rho}(\mbf{r})\hat{n}(\mbf{r}^\prime)\rangle_\phi -
       \langle\hat{\rho}(\mbf{r})\rangle_\phi\langle\hat{n}(\mbf{r}^\prime)\rangle_\phi\big].
\end{equation}
which is the covariance of the charge density with the number
density. Unlike $\chi^{(n)}_n$, however,
$\chi^{(\rho)}_{n}$ is not
directly related to derivatives of $\mcl{F}^{\rm (ex)}_{\rm
intr}[\varrho,\beta\varphi]$. To make the connection explicit,
we therefore also consider the hyperdirect
correlation functionals defined by the mixed functional derivatives
\begin{align}
 \label{eqn:cn-rhophi}
    c^{(\varrho\varphi)}_{n}(\mbf{r}, \mbf{r}^\prime; [\varrho,\beta\varphi]) 
    =
    \frac{\delta^2 \beta \mcl{F}_{\mrm{intr}}^{\mrm{(ex)}}[\varrho, \beta\varphi]}{\delta\beta\varphi(\mbf{r}^\prime)\delta\varrho(\mbf{r})}
    =- \frac{\delta c^{(1)}(\mbf{r}; [\varrho, \beta\varphi])}{\delta \beta\varphi(\mbf{r}^\prime)} 
\end{align}
and
\begin{equation}
 \label{eqn:cn-phirho}
    c^{(\varphi\varrho)}_{n}(\mbf{r}, \mbf{r}^\prime; [\varrho,\beta\varphi]) =
    \frac{\delta^2 \beta \mcl{F}_{\mrm{intr}}^{\mrm{(ex)}}[\varrho, \beta\varphi]}{ \delta\varrho(\mbf{r}^\prime)\delta\beta\varphi(\mbf{r})}
= \frac{\delta n^{(1)}(\mbf{r}; [\varrho, \beta\varphi])}{\delta \varrho(\mbf{r}^\prime)}.
\end{equation}
For the second equalities in Eqs.~\ref{eqn:cn-rhophi}
and \ref{eqn:cn-phirho}, we have used the definitions provided
by Eqs.~\ref{eqn:DCF-1} and \ref{eqn:n-Fex}, respectively.
Assuming $\mcl{F}_{\mrm{intr}}^{\mrm{(ex)}}[\varrho, \beta\varphi]$ is
continuous such that Schwarz's theorem holds, the relationship between
these two hyperdirect correlation functionals is
\begin{equation}
\label{eqn:schwarz}
c^{(\varrho\varphi)}_{n}(\mbf{r}, \mbf{r}^\prime; [\varrho,\beta\varphi]) = c^{(\varphi\varrho)}_{n}(\mbf{r}^\prime, \mbf{r}; [\varrho,\beta\varphi]).
\end{equation}

Functional differentiation of the Euler--Lagrange equation
(Eq.~\ref{eqn:Euler-Lagrange-full}) with respect to
$\beta\varphi(\mbf{r}^\prime)$, followed by substitution from
Eqs.~\ref{eqn:DCF-2} and~\ref{eqn:cn-rhophi},
relates $c^{(\varrho\varphi)}_{n}$,
$\chi^{(\rho)}_{n}$ and $c^{(2)}$ at equilibrium
via an exact hyperdensity OZ equation
\begin{equation}
\label{eqn:hyperOZ}
\begin{split}
&c^{(\varrho\varphi)}_{n} (\mbf{r},\mbf{r}^\prime; [\rho,\phi]) =  \\
&\frac{\chi^{(\rho)}_n(\mbf{r}, \mbf{r}^\prime)}{\rho(\mbf{r})} 
 - \int\!\!\mrm{d}\mbf{r}^{\prime\prime}\, c^{(2)}(\mbf{r},\mbf{r}^{\prime\prime};[\rho,\phi])\,\chi^{(\rho)}_{n}(\mbf{r}^{\prime\prime},\mbf{r}^\prime).
\end{split}
\end{equation}
%

\subsection{Response to an external non-electrostatic potential}

We can also consider the response to the external (non-electrostatic) potential
$V_{\rm ext}$. As in regular cDFT, 
the ``number--number'' response function is
\begin{equation}
\label{eqn:chirhorho}
\chi^{(\rho)}_\rho(\mbf{r},\mbf{r}^\prime) = -\left.\frac{\delta\rho(\mbf{r})}{\delta\beta V_{\rm ext}(\mbf{r}^\prime)}\right\vert_{\beta\phi,\, \beta\mu},
\end{equation}
which can be readily shown to be the variance of the local number
density,
\begin{equation}
 \chi^{(\rho)}_\rho(\mbf{r},\mbf{r}^\prime) =
  \big[\langle\hat{\rho}(\mbf{r})\hat{\rho}(\mbf{r}^\prime)\rangle_\phi -
       \langle\hat{\rho}(\mbf{r})\rangle_\phi\langle\hat{\rho}(\mbf{r}^\prime)\rangle_\phi\big].
\end{equation}
Whereas $\chi^{(\rho)}_\rho$ can be considered a ``direct'' response,
any changes to the charge density  only occur ``indirectly'' through
changes in the number density.
Defining the ``charge--number'' response function,
\begin{equation}
\label{eqn:chi-nrho-def}
\left.\chi^{(n)}_\rho(\mbf{r},\mbf{r}^\prime) = -\frac{\delta n(\mbf{r})}{\delta \beta V_{\mrm{ext}}(\mbf{r}^\prime)}\right\vert_{\beta\phi,\, \mu, T},
\end{equation}
applying the chain rule, and using Eqs.~\ref{eqn:cn-phirho}
and~\ref{eqn:chirhorho}, we find
\begin{equation}
\label{eqn:chi-nrho}
\begin{split}
\chi^{(n)}_\rho(\mbf{r},\mbf{r}^\prime) = 
\int\!\mrm{d}\mbf{r}^{\prime\prime}\,c^{(\varphi\varrho)}_n(\mbf{r},\mbf{r}^{\prime\prime};[\rho, \beta\phi])\,
\chi^{(\rho)}_\rho(\mbf{r}^{\prime\prime},\mbf{r}^\prime).
\end{split}
\end{equation}
Here, we have limited our discussion to fluctuation profiles directly
relevant to electromechanical coupling, which we summarize in
Fig.~\ref{matrix}.  We note that measures of the number and charge
density fluctuations due to change in the chemical potential and
temperature can also be considered, e.g.,
$\chi^{(\rho)}_{\mu}, \chi^{(n)}_{\mu}, \chi^{(\rho)}_{T}$ and
$ \chi^{(n)}_{T}$. These can be particularly insightful near a phase
transition
\cite{Evans2015, Eckert2020, Coe2022, Eckert2023, Bui2024}.


\subsection{Discussion}

Using hyper-DFT, we have provided a rigorous statistical mechanical
framework in which to understand the response of a fluid to both
changes in the external non-electrostatic potential, and the external
electrostatic potential. Our expressions for the direct
correlation functionals (Eq.~\ref{eqn:DCF-alpha}) and number--number
response function $\chi^{(\rho)}_\rho$ (Eq.~\ref{eqn:chirhorho}) are
similar to those readily obtained from cDFT, but generalized to cases
with an external electrostatic potential. For $\chi^{(n)}_n$, while
its microscopic interpretation at equilibrium as the variance of the
charge density (Eq.~\ref{eqn:chi-nn-var}) is well known
\cite{HansenMcDonaldBook, Chandler1977, Rodgers2009}, its definition
as a hyperfunctional (Eq.~\ref{eqn:chi-nn-def}) gives it deeper
significance, and holds for general $(\varrho, \varphi)$. Moreover,
the hyperdirect correlation functionals $c^{(\varrho\varphi)}_n$ and
$c^{(\varphi\varrho)}_n$ (Eqs.~\ref{eqn:cn-rhophi}
and~\ref{eqn:cn-phirho}) are only defined within the hyper-DFT
framework. As we see from Eqs.~\ref{eqn:hyperOZ}
and~\ref{eqn:chi-nrho}, these hyperdirect correlation functionals play
an integral role in describing the electromechanical response of a
fluid. Hyper-DFT is the natural framework in which to
understand electromechanics from a microscopic perspective.

We emphasize that we have introduced a set of equations that allow us
to sensibly discuss the charge density $n(\mbf{r})$, without resort to
any knowledge of the underlying intermolecular interactions. The
framework we present therefore applies equally well to conductors
(e.g., electrolytes) or dielectric media (e.g., polar fluids), which
is a significant advantage for discussing their electromechanical
response on the same footing. For example, the number--charge response 
function
$\chi^{(\rho)}_n$ has mostly been discussed solely in the context
of ``rigid ion'' models \cite{HansenMcDonaldBook}. In the hyper-DFT
framework, electrostrictive response in polar fluids is also measured
through $\chi^{(\rho)}_n$, without the need to resolve molecular
orientations.

As shown in Refs.~\cite{Sammuller2024hdft,Sammuller2024review}, the
hyper-DFT framework also lends itself naturally to a machine learning
(ML) procedure dubbed ``neural functional
theory'' \cite{Sammuller2023}, making it amenable to practical
computation. Extending such an approach in the present case would
translate to using data from grand canonical simulations to
learn the first derivatives of $\mcl{F}^{\mrm{(ex)}}_{\rm
intr}[\varrho, \beta\varphi]$, i.e.,
$c^{(1)}(\mbf{r};[\varrho, \beta\varphi])$ and
$n^{(1)}(\mbf{r};[\varrho, \beta\varphi])$.
However, the success of
the neural functional approach relies upon the locality of
correlations, which is generally not a safe assumption when
considering electromechanical phenomena. Recently, we showed how LMFT
can be used with such a ML procedure to circumvent this issue in the
case of primitive models for ionic
fluids \cite{bui2024learningclassicaldensityfunctionals}, exploiting
the connection between cDFT and LMFT established by Archer and
Evans \cite{Archer2013}.  A similar tactic would prove useful in this
context.  In the following section, we therefore establish the
connection between hyper-DFT and LMFT.

\section{The relationship between LMFT and hyper-DFT} \label{ref:LMFTconnection}
\label{sec:LMFT}

LMFT is a statistical mechanical framework that aims to recast a
system with long-ranged interactions in terms of a reference system
with short-ranged interactions, which, in the presence of an
appropriate one-body potential $\phi_{\rm R}(\mbf{r})$, recovers the
same one-body density \cite{Rodgers2008}.
The reference system together with $\phi_{\rm
R}$ is dubbed the ``mimic'' system. Indicating properties of the mimic
system with the ``${\rm R}$'' subscript, the LMF condition states
\begin{equation}
  \label{eqn:LMF-1}
 \langle \hat{\rho}(\mbf{r}, \{\mbf{r}_i\})\rangle_{\phi} = \langle\hat{\rho}_{\rm R}(\mbf{r}, \{\mbf{r}_i\})\rangle_{\phi_{\rm R}}.
\end{equation}
In cases where LMFT is applied to electrostatic interactions, we also
require a second LMF condition,
\begin{equation}
  \label{eqn:LMF-2}
\langle \hat{n}(\mbf{r},\mbf{R}^N)\rangle_\phi = \langle \hat{n}_{\rm R}(\mbf{r},\mbf{R}^N)\rangle_{\phi_{\rm R}}.
\end{equation}

The original derivations of LMFT are based on the Yvon--Born--Green
(YBG) hierarchy of equations \cite{HansenMcDonaldBook}.
When applied to electrostatic
interactions, Weeks and coworkers 
\cite{Rodgers2008,Rodgers2008b,Weeks1998,Rodgers2006,Chen2004,Remsing2016, Gao2020}
have extensively demonstrated that the LMF conditions can be met with,
\begin{equation}
 \label{eqn:phiR-weeks}
  \phi_{\mrm{R}}(\mbf{r}) = \phi(\mbf{r})
  + \int\!\!\mrm{d}\mbf{r}^{\prime}\,n_{\mrm{R}}(\mbf{r}^{\prime}) v_1(|\mbf{r}-\mbf{r}^{\prime}|),
\end{equation}
where
\begin{equation}
 v_1(r) =  \frac{\erf(\kappa r)}{r},
\end{equation}
with $\kappa^{-1}$ defining a range separation of the Coulomb
potential,
\begin{equation}
    \frac{1}{r} = v_0(r) + v_1(r). 
    \label{eqn:splitting}
\end{equation}
In the mimic system, electrostatic interactions are replaced by their
short-ranged counterpart, such that the potential energy reads
\begin{equation}
\begin{split}
    \hat{\mcl{U}}_{\mrm{R}}(\mbf{R}^N) &=  \hat{\mcl{U}}_{\mrm{ne}}(\mbf{R}^N) \\
    & +  \frac{1}{2} \int\!\!\mrm{d}\mbf{r}\!\!\int\!\!\mrm{d}\mbf{r}^{\prime}\,\hat{n}_{\mrm{R}}(\mbf{r},\mbf{R}^N)\hat{n}_{\mrm{R}}(\mbf{r}^{\prime},\mbf{R}^N)v_0(|\mbf{r}-\mbf{r}^{\prime}|).
\end{split}
\end{equation}

The LMFT conditions (Eqs.~\ref{eqn:LMF-1} and~\ref{eqn:LMF-2}) are
based on structure. In the language of hyper-DFT, they pertain to the
first functional derivatives of $\mcl{F}^{\rm (ex)}_{\rm
intr}[\varrho,\beta\varphi]$: in the case of the first LMF condition
(Eq.~\ref{eqn:LMF-1}), this is via the Euler--Lagrange equation
(Eq.~\ref{eqn:Euler-Lagrange-full}); for the second LMF condition one
can see this directly from Eq.~\ref{eqn:n-Fex}. The YBG equations upon
which LMFT is based relate these structural properties to two-body
correlations. In the language of hyper-DFT, these pertain to second
derivatives of $\mcl{F}^{\rm (ex)}_{\rm intr}[\varrho,\beta\varphi]$.

In a density functional approach, whether cDFT or hyper-DFT, it is
natural to consider the reference system at the level of $\mcl{F}^{\rm
(ex)}_{\rm intr}[\varrho,\beta\varphi]$ itself, i.e.,
\begin{equation}
 \label{eqn:Fex-loc-nonloc}
 \mcl{F}^{\rm (ex)}_{\rm intr}[\varrho,\beta\varphi] = \mcl{F}^{\rm (ex)}_{\rm intr,R}[\varrho,\beta\varphi] + \Delta\mcl{F}^{\rm (ex)}_{\rm intr}[\varrho,\beta\varphi].
\end{equation}
Here, $\mcl{F}^{\rm (ex)}_{\rm intr,R}$ is defined as the local part
of the free energy functional of the system such that all non-local
contributions are contained within $\Delta\mcl{F}^{\rm (ex)}_{\rm
intr}$. However, when considered on its own, $\mcl{F}^{\rm (ex)}_{\rm intr,R}$ is
simply a free energy functional for a system whose correlations are
local. One such reference system could be that in which electrostatic
interactions are described by the short-ranged potential $v_0$. This
reference system would then satisfy its own Euler--Lagrange equation
\begin{equation}
  \rho_{\rm R}(\mbf{r}) =\frac{\tcr{\zeta}}{\Lambda^3}
  \exp\left(-\beta [V_{\rm ext}(\mbf{r}) - \mu_{\rm R}] + c_{\rm R}^{(1)}(\mbf{r};[\rho_{\rm R},\beta\phi_{\rm R}]) \right).
  \label{eqn:Euler-Lagrange-mimic}
\end{equation}
Note that, as the reference system only differs from the true system in its electrostatic interactions, 
it feels the same non-electrostatic potential $V_{\rm ext}$. Subtracting Eq.~\ref{eqn:Euler-Lagrange-mimic} from
Eq.~\ref{eqn:Euler-Lagrange-full}, enforcing the first LMF condition
(Eq.~\ref{eqn:LMF-1}), and rearranging gives
\begin{equation}
 \label{eqn:c1-diff}
 c^{(1)}(\mbf{r};[\rho^{\rm },\beta\phi]) - c_{\rm R}^{(1)}(\mbf{r};[\rho_{\rm R},\beta\phi_{\rm R}]) =
 -\beta\Delta\mu,
\end{equation}
where $\Delta\mu = \mu - \mu_{\rm R}$. Similarly, in the case of the
charge densities, for the mimic system we have
\begin{equation}
\label{eqn:nR-Fex}
n^{(1)}_{\rm R}(\mbf{r};[\rho_{\rm R},\beta\phi_{\rm R}]) =
\left.\frac{\delta \mcl{F}^{(\rm ex)}_{\rm intr, R}[\rho_{\rm R},\beta\varphi]}{\delta\beta\varphi(\mbf{r})}\right\vert_{\varphi=\phi_{\rm R}}.
\end{equation}
Enforcing the second LMF condition (Eq.~\ref{eqn:LMF-2}) we obtain
\begin{equation}
 \label{eqn:n-nR-Fex}
n^{(1)}(\mbf{r};[\rho,\beta\phi]) = n^{(1)}_{\rm R}(\mbf{r};[\rho,\beta\phi_{\rm R}]).
 \end{equation}

At a fundamental level,
the strategy in LMFT is to find $\phi_{\rm R}$ that satisfies
the LMF conditions. In a density functional approach, one instead aims
to find $\Delta\mcl{F}^{\rm (ex)}_{\rm intr}$ whose first derivatives
satisfy the LMF conditions. In the case of cDFT, Archer and Evans have
shown that the two approaches are equivalent when LMFT is applied to
non-electrostatic interactions (i.e., Eqs.~\ref{eqn:LMF-1}
and~\ref{eqn:c1-diff}) \cite{Archer2013}.  In the present case, we
also need to consider the second LMF condition (i.e.,
Eqs.~\ref{eqn:LMF-2} and~\ref{eqn:n-nR-Fex}). The advantage of the YBG
formulation of LMFT is that conditions that the reference system
should obey are made explicit; we refer the reader to
Ref.~\cite{Rodgers2008} for details. The advantage of the density
functional formalism is direct access to free energies and
thermodynamic consistency. In the case of the hyper-DFT formalism
that follows, we recover information on the conditions that are placed
on the reference system.

\subsection{An explicit, yet exact, expression for $\phi_{\rm R}$ in hyper-DFT}

Equations~\ref{eqn:c1-diff} and~\ref{eqn:n-nR-Fex} are exact but do
not instruct on how to prescribe $\phi_{\rm R}$ (or equivalently,
$\Delta\mcl{F}^{\rm (ex)}_{\rm intr}$). Inspecting $\phi_{\rm R}$ as
derived by Weeks and co-workers (Eq.~\ref{eqn:phiR-weeks}) strongly
suggests that an appropriate $\Delta\mcl{F}^{\rm (ex)}_{\rm intr}$
will be of a mean field form, as indeed is the case in cDFT. Before
making any approximations, however, it is instructive to obtain an
explicit relationship between $\Delta\mcl{F}^{\rm (ex)}_{\rm intr}$
(or more precisely, its first functional derivatives) and $\phi_{\rm
R}$ that remains exact.

As we know that the LMF conditions pertain to first functional
derivatives of the excess Helmholtz free energy functional, we
start by differentiating Eq.~\ref{eqn:Fex-loc-nonloc},
\begin{subequations}
\label{eqn:delFex-loc-nonloc}
\begin{align}
 \frac{\delta \mcl{F}^{(\rm ex)}_{\rm intr}[\varrho,\beta\varphi]}{\delta\varrho(\mbf{r})} &=
 \frac{\delta \mcl{F}^{(\rm ex)}_{\rm intr,R}[\varrho,\beta\varphi]}{\delta\varrho(\mbf{r})} +
 \frac{\delta \Delta\mcl{F}^{(\rm ex)}_{\rm intr}[\varrho,\beta\varphi]}{\delta\varrho(\mbf{r})}, \label{eqn:delFex-loc-nonloc-rho} \\
 \frac{\delta \mcl{F}^{(\rm ex)}_{\rm intr}[\varrho,\beta\varphi]}{\delta\beta\varphi(\mbf{r})} &=
 \frac{\delta \mcl{F}^{(\rm ex)}_{\rm intr,R}[\varrho,\beta\varphi]}{\delta\beta\varphi(\mbf{r})} +
 \frac{\delta \Delta\mcl{F}^{(\rm ex)}_{\rm intr}[\varrho,\beta\varphi]}{\delta\beta\varphi(\mbf{r})}. \label{eqn:delFex-loc-nonloc-phi}
\end{align}
\end{subequations}
Motivated by the observation that Eq.~\ref{eqn:phiR-weeks} can be
written in the form 
\begin{equation}
 \phi_{\rm R}(\mbf{r}) = \phi(\mbf{r}) - \Delta\phi(\mbf{r}),
\end{equation}
we perform functional series expansions of both $\delta \mcl{F}^{(\rm
ex)}_{\rm intr,R}[\varrho,\beta\varphi]/\delta\varrho(\mbf{r})$ and
$\delta \mcl{F}^{(\rm ex)}_{\rm
intr,R}[\varrho,\beta\varphi]/\delta\beta\varphi(\mbf{r})$ in
$\beta\varphi$. Recalling earlier definitions for
$c_n^{(\varrho\varphi)}$ (Eq.~\ref{eqn:cn-rhophi}) and $\chi_n^{(n)}$
(Eq.~\ref{eqn:chi-nn-Fex}), we have
\begin{widetext}
\begin{subequations}
\begin{align}
\frac{\delta \mcl{F}^{(\rm ex)}_{\rm intr,R}[\varrho,\beta\varphi]}{\delta\varrho(\mbf{r})} =
\frac{\delta \mcl{F}^{(\rm ex)}_{\rm intr,R}[\varrho,\beta\varphi-\beta\Delta\varphi]}{\delta\varrho(\mbf{r})}  +
\int\!\mrm{d}\mbf{r}^\prime\,c_{n,\mrm{R}}^{(\varrho\varphi)}(\mbf{r},\mbf{r}^\prime;[\varrho,\beta\varphi])\Delta\varphi(\mbf{r}^\prime) +
\mathcal{O}\big(\Delta\varphi^2\big), \\
\frac{\delta \mcl{F}^{(\rm ex)}_{\rm intr,R}[\varrho,\beta\varphi]}{\delta\beta\varphi(\mbf{r})} =
\frac{\delta \mcl{F}^{(\rm ex)}_{\rm intr,R}[\varrho,\beta\varphi-\beta\Delta\varphi]}{\delta\beta\varphi(\mbf{r})} -
\int\!\mrm{d}\mbf{r}^\prime\,\chi_{n,\mrm{R}}^{(n)}(\mbf{r},\mbf{r}^\prime;[\varrho,\beta\varphi])\Delta\varphi(\mbf{r}^\prime) +
\mathcal{O}\big(\Delta\varphi^2\big).
\end{align}
\end{subequations}
\end{widetext}
Upon substitution of 
Eqs.~\ref{eqn:delFex-loc-nonloc} into 
the above expansions, and
recognizing the definitions
of the one-body direct correlation functional (Eq.~\ref{eqn:DCF-1})
and one-body charge density functional (Eq.~\ref{eqn:n-Fex}), we
arrive at the following relationships
\begin{widetext}
\begin{subequations}
\label{eqn:Fex-expand2}
\begin{align}
-c^{(1)}(\mbf{r};[\varrho,\beta\varphi]) -\frac{\delta\beta\Delta\mcl{F}^{(\rm ex)}_{\rm intr}[\varrho,\beta\varphi]}{\delta\varrho(\mbf{r})} &=  -c_{\rm R}^{(1)}(\mbf{r};[\varrho,\beta\varphi-\beta\Delta\varphi])
 +
\beta\int\!\mrm{d}\mbf{r}^\prime\,c_{n,\mrm{R}}^{(\varrho\varphi)}(\mbf{r},\mbf{r}^\prime;[\varrho,\beta\varphi])\Delta\varphi(\mbf{r}^\prime) +
\mathcal{O}\big(\Delta\varphi^2\big), \label{eqn:Fex-expand2-rho} \\
n^{(1)}(\mbf{r};[\varrho,\beta\varphi]) -\frac{\delta\beta\Delta\mcl{F}^{(\rm ex)}_{\rm intr}[\varrho,\beta\varphi]}{\delta\beta\varphi(\mbf{r})} &= n^{(1)}_{\rm R}(\mbf{r};[\varrho,\beta\varphi-\beta\Delta\varphi]) 
 -
\beta\int\!\mrm{d}\mbf{r}^\prime\,\chi_{n,\mrm{R}}^{(n)}(\mbf{r},\mbf{r}^\prime;[\varrho,\beta\varphi])\Delta\varphi(\mbf{r}^\prime) +
\mathcal{O}\big(\Delta\varphi^2\big). \label{eqn:Fex-expand2-phi}
\end{align}
\end{subequations}
\end{widetext}
%
%
By setting $\varphi = \phi$, $\Delta\varphi = \phi-\phi_{\rm
R} \equiv \Delta\phi$, evaluating Eqs.~\ref{eqn:Fex-expand2-rho}
and~\ref{eqn:Fex-expand2-phi} at equilibrium, and evoking the LMF conditions,
%
we obtain the following exact expressions
\begin{widetext}
\begin{subequations}
\label{eqn:Fex-expand-final}
\begin{align}
\left.\frac{\delta\beta\Delta\mcl{F}^{(\rm ex)}_{\rm intr}[\varrho,\beta\phi]}{\delta\varrho(\mbf{r})}\right\vert_{\varrho=\rho} &=
\beta\Delta\mu - \beta\int\!\mrm{d}\mbf{r}^\prime\,c_{n,\mrm{R}}^{(\varrho\varphi)}(\mbf{r},\mbf{r}^\prime;[\rho ,\beta\phi])\Delta\phi(\mbf{r}^\prime)
+ \mathcal{O}\big(\Delta\phi^2\big), \label{eqn:Fex-expand-final-rho} \\
\left.\frac{\delta\beta\Delta\mcl{F}^{(\rm ex)}_{\rm intr}[\rho,\beta\varphi]}{\delta\beta\varphi(\mbf{r})}\right\vert_{\varphi=\phi} &= 
\beta\int\!\mrm{d}\mbf{r}^\prime\,\chi_{n,\mrm{R}}^{(n)}(\mbf{r},\mbf{r}^\prime;[\rho,\beta\phi])\Delta\phi(\mbf{r}^\prime)
+ \mathcal{O}\big(\Delta\phi^2\big). \label{eqn:Fex-expand-final-phi}
\end{align}
\end{subequations}
\end{widetext}

\subsection{Asserting a mean field form for $\Delta\mcl{F}^{\rm (ex)}_{\rm intr}$}

Provided that $\phi_{\rm R}$ exists,
Eqs.~\ref{eqn:Fex-expand-final-rho} and~\ref{eqn:Fex-expand-final-phi}
specify exactly how $\phi_{\rm R} = \phi-\Delta\phi$ and
$\Delta\mcl{F}^{\rm (ex)}_{\rm intr}$ are related, but they are still
of little use unless $\Delta\mcl{F}^{\rm (ex)}_{\rm intr}$ is
specified. One way forward is make the following mean field
approximation
\begin{widetext}
\begin{minipage}{\textwidth}
\begin{equation}
\label{eqn:DeltaFex-MF}
 \Delta\mcl{F}^{\rm (ex)}_{\rm intr}[\varrho,\beta\varphi] =
\Delta\mu\int\!\mrm{d}\mbf{r}^\prime\,\varrho(\mbf{r}^\prime) +
  \frac{1}{2}\int\!\mrm{d}\mbf{r}^\prime\!\int\!\mrm{d}\mbf{r}^{\prime\prime}\,
  n^{(1)}_{\rm R}(\mbf{r}^\prime;[\varrho,\beta\varphi])n^{(1)}_{\rm R}(\mbf{r}^{\prime\prime};[\varrho,\beta\varphi])v_1(|\mbf{r}^\prime-\mbf{r}^{\prime\prime}|).
\end{equation}
\end{minipage}
\end{widetext}
Taking the first functional derivatives, 
noting Eqs.~\ref{eqn:chi-nn-def}, \ref{eqn:cn-phirho}, and~\ref{eqn:schwarz},
setting $\varphi=\phi$, and evaluating at equilibrium, gives
\begin{widetext}
\begin{subequations}
\label{eqn:Fex-MF-final}
\begin{align}
\left.\frac{\delta\beta\Delta\mcl{F}^{(\rm ex)}_{\rm intr}[\varrho,\beta\phi]}{\delta\varrho(\mbf{r})}\right\vert_{\varrho=\rho} &= 
\beta\Delta\mu +\beta\int\!\mrm{d}\mbf{r}^\prime\!\int\!\mrm{d}\mbf{r}^{\prime\prime}\,c_{n,\mrm{R}}^{(\varrho\varphi)}(\mbf{r},\mbf{r}^\prime;[\rho,\beta\phi])
n^{(1)}_{\rm R}(\mbf{r}^{\prime\prime};[\rho,\beta\phi])v_1(|\mbf{r}^\prime-\mbf{r}^{\prime\prime}|), \label{eqn:Fex-MF-final-rho} \\
\left.\frac{\delta\beta\Delta\mcl{F}^{(\rm ex)}_{\rm intr}[\rho,\beta\varphi]}{\delta\beta\varphi(\mbf{r})}\right\vert_{\varphi=\phi} &=-
\beta\int\!\mrm{d}\mbf{r}^\prime\!\int\!\mrm{d}\mbf{r}^{\prime\prime}\,\chi_{n,\mrm{R}}^{(n)}(\mbf{r},\mbf{r}^\prime;[\rho,\beta\phi])
n^{(1)}_{\rm R}(\mbf{r}^{\prime\prime};[\rho,\beta\phi])v_1(|\mbf{r}^\prime-\mbf{r}^{\prime\prime}|). \label{eqn:Fex-MF-final-phi}
\end{align}
\end{subequations}
\end{widetext}
Ignoring terms $\mathcal{O}(\Delta\phi^2)$ and higher, when comparing
Eqs.~\ref{eqn:Fex-expand-final} and~\ref{eqn:Fex-MF-final}, one can
deduce that
\begin{equation}
\label{eqn:DeltaPhi}
\Delta\phi(\mbf{r}^\prime) = -\int\!\!\mrm{d}\mbf{r}^{\prime\prime}\,n^{(1)}_{\mrm{R}}(\mbf{r}^{\prime\prime};[\rho,\beta\phi]) v_1(|\mbf{r}^{\prime}-\mbf{r}^{\prime\prime}|).
\end{equation}
This is the same expression that Weeks and co-workers have derived (Eq.~\ref{eqn:phiR-weeks})
based on the YBG hierarchy. Provided that the reference system is
chosen such that we can ignore the higher order terms in
Eqs.~\ref{eqn:Fex-expand-final-rho}
and~\ref{eqn:Fex-expand-final-phi}, LMFT is the same as hyper-DFT in
the mean field approximation specified by Eq.~\ref{eqn:DeltaFex-MF}.

\subsection{Discussion}

Our derivation above bears some obvious similarities to that of Archer
and Evans \cite{Archer2013}, who demonstrated a relationship between
LMFT and cDFT, but also some key differences. It is worth
comparing certain aspects of the two approaches. First, Archer and
Evans did not consider electrostatic interactions and, therefore, it
is unclear whether the insights into the relationship between LMFT and
cDFT obtained in Ref.~\cite{Archer2013} apply to
electromechanical systems in general. An important exception is the
class of models whose charge density operator takes the form
\begin{equation}
\label{eqn:PM-nop}
\hat{n}(\mbf{r}) = \sum_\gamma  q_{\gamma} \hat{\rho}_{\gamma}(\mbf{r}) ,
\end{equation}
%
where $\gamma$ labels a particle's type. 
Note that this is not a
sum over interaction sites; recall that a particle is defined as a
body whose chemical potential is specified by the reservoir. In such
cases, the functional dependence between the charge and number
densities is trivial,
and the grand potential functional
in Eq.~\ref{eqn:OmegaRhoPhi} reduces to
\begin{equation}
\varOmega[\varrho] = \mcl{F}^{\mrm{(id)}}_{\mrm{intr}}[\varrho] + \mcl{F}^{\mrm{(ex)}}_{\mrm{intr}}[\varrho] + \int\!\!\mrm{d}\mbf{r}\, \varrho(\mbf{r}) [q \varphi(\mbf{r}) + V_{\rm ext}(\mbf{r})- \mu],
\end{equation}
i.e., the charge density can be written explicitly in terms of the number density,
and the results from Ref.~\cite{Archer2013} can be used directly.  We
recently exploited this fact to use ML to accurately represent the
one-body direct correlation functionals for primitive models of ionic
fluids \cite{bui2024learningclassicaldensityfunctionals}. In this
work, by connecting hyper-DFT and LMFT, we open the possibility of
applying similar ML strategies to scenarios where the functional
dependence is \emph{a priori} unknown, e.g., polar molecules.

Second, as the original derivations of LMFT by Weeks and co-workers
are based on the YBG hierarchy, they provide important insights
regarding the construction of the mean field approximation. We refer
the reader to Ref.~\cite{Rodgers2008} for a detailed discussion, but
in simple terms, it amounts to retaining enough of the interaction
potential in the mimic system; in the present case, the means choosing
$\kappa^{-1}$ sufficiently large to capture enough correlations
explicitly in $\mcl{F}_{\rm intr,R}^{\rm (ex)}$. A criticism
raised in Ref.~\cite{Remsing2016} is that this well-controlled nature
of the mean field approximation is lost when constructing the
mean field approximation at the level of $\Delta \mcl{F}_{\rm
intr,R}^{\rm (ex)}$.  In the hyper-DFT formalism, ignoring terms
$\mathcal{O}(\Delta\phi^2)$ and higher in
Eqs.~\ref{eqn:Fex-expand-final-rho} and~\ref{eqn:Fex-expand-final-phi}
is tantamount to choosing $\kappa^{-1}$ large enough that both
$c_{n,\mrm{R}}^{(\varrho\varphi)}(\mbf{r},\mbf{r}^\prime;[\rho,\beta\phi])$
and $\chi^{(n)}_{n,\mrm{R}}(\mbf{r},\mbf{r}^\prime;[\rho,\beta\phi])$
ensure that this truncation at first order is accurate. It is highly
likely that
$\chi^{(n)}_{n,\mrm{R}}(\mbf{r},\mbf{r}^\prime;[\rho,\beta\phi])$
determines the minimum value of $\kappa^{-1}$ that can be used. (For
example, for the primitive models specified by Eq.~\ref{eqn:PM-nop},
$c_{n,\gamma}^{(\varrho_{\gamma}\varphi)}(\mbf{r},\mbf{r}^\prime;[\{\varrho_{\gamma}\},\beta\varphi])
=
c_{n,\gamma}^{(\varphi\varrho_{\gamma})}(\mbf{r}^\prime,\mbf{r};[\{\varrho_{\gamma}\},\beta\varphi])$
is strictly local.) This statement therefore complements, rather than
contradicts, the derivation of Archer and Evans. In
Appendix \ref{sec:appendix-nonelectro}, we also recast the
relationship between LMFT and hyper-DFT relationship for the
non-electrostatic case considered in Ref.~\cite{Archer2013}.

Our final remark on the comparison to Ref.~\cite{Archer2013} is a subtle one,
and concerns the difference in chemical potentials between the true
and mimic systems. Presumably to make the connection to LMFT (as
derived by from the YBG hierarchy) as explicit as possible, Archer and
Evans took the spatial gradient of the Euler--Lagrange equations. Upon
integrating, $\Delta\mu$ was then obtained by identifying it as the
integration constant. (This is also the case in the original LMFT
derivations.) In contrast, in our approach the conditions placed on
the reference system (i.e., choosing $\kappa^{-1}$ sufficiently large)
will imply that the difference in bulk free energies between the true
and mimic systems will be dominated by differences in potential energy
rather than entropy. This fact facilitates the derivation of
expressions for $\Delta\mu$ in terms of bulk susceptibilities, which
is particularly advantageous when considering electrostatic
interactions. Expressions for $\Delta\mu$ will depend on the system
under investigation. In Appendix~\ref{sec:polarmu} we
give the expression for a polar fluid, while in Appendix~\ref{sec:dielectricconstant} we discuss 
how $\chi_{n,\mrm{R}}^{(n)}$ is related to the dielectric constant 
of the system.

\section{Conclusions and Outlook} \label{sec:conclusions}

In this work, we have extended the hyper-DFT approach presented by
Samm\"{u}ller \etal{} \cite{Sammuller2024hdft, Sammuller2024review} to derive rigorous
statistical mechanical expressions for various response and
(hyper)direct correlation functions (or functionals) relevant to
electromechanical phenomena in fluids. In terms of the theoretical
structure, a difference from the original hyper-DFT articles is that
we have considered a spatially-varying collective variable (i.e., the
charge density) as our observable. While this results in some obvious
superficial consequences, a deeper repercussion is that this has led
us to extend the definitions of the hyperdirect correlation
functionals in terms of mixed second derivatives of the excess
intrinsic Helmholtz free energy functional.  But more significant
from a theoretical perspective is that we have applied hyper-DFT to a
scenario in which the observable itself contributes to the
interparticle potential energy of the system.

A pleasing aspect of this hyper-DFT approach is that it naturally
captures the both the charge and number density responses that arise
not only as a ``direct'' consequence of an external perturbation, but
also ``indirectly'' due to correlations between the number and charge
densities, without the added complexity of needing to resolve
orientations \cite{Teixeira1991,Frodl1992,Jeanmairet2013,Simon2025,yang2024}
or intramolecular interaction sites \cite{Chandler1986,
Sundararaman2017, Chuev2021}.  In a bid to go beyond cDFT for the
one-body density, one might naturally be inclined to introduce
functional dependence of the free energy on 
the charge density or
polarization fields (see, e.g., Refs.~\onlinecite{Jeanmairet2013,
Jeanmairet2016}).
In such cases, electromechanical coupling needs
to be introduced at a constitutive level, possibly
with parameterization from molecular simulations. In
contrast, in the hyper-DFT framework where the electrostatic potential
enters the free energy functional, electromechanics is an
emergent phenomenon.

This emergent nature of electromechanics in the theory presented will
be powerful for understanding how fluids respond to spatially-varying
electric fields. In this context, ``respond'' can be taken to mean not
only the changes in fluid behavior as described directly by the
response functions discussed in this work, but also how other complex
emergent phenomena, e.g., phase behavior, are impacted. The strategy
is a simple one: specify the external potentials, and solve the
Euler--Lagrange equation. The powerful nature of this approach should
be made practical by the connection between hyper-DFT and local
molecular field theory that we have established.  We refer the readers
to Ref.~\cite{bui2025efg} where we have applied this hyper-DFT
framework to explore ``dielectrocapillarity,'' i.e., how capillarity of dielectric fluids is controlled by
electric field gradients.  We speculate that the theoretical
foundations presented here will inform, and augment, modelling
methods for electromechanics, such as lattice-based
techniques \cite{Ruiz-Gutierrez2021}, density functional methods
applied away from equilibrium \cite{Illien2024}, and continuum
mechanics-based approaches \cite{Sprik2021}.


One obvious question to pose is whether the framework that we have
presented can be used to probe the behavior  of the 
dielectric constant of fluids confined at small length scales. 
This topic has a 
long history that has also garnered significant attention
\cite{Schlaich2016, Cox2022, Olivieri2021, Matyushov2021, Borgis2023}
more recently owing to significant
advances in fabricating confined environments at the nanoscale
\cite{Geim2013, Fumagalli2018}.
However, one of the main motivations to formulate phenomena in terms
of continuum-style quantities such as the dielectric constant is
to describe a system's behavior using fewer degrees of
freedom. When done appropriately, such an endeavor can not only reduce
the computational complexity of a problem, but also aid conceptual
understanding. The hyper-DFT approach to electromechanics that we have
presented should achieve both of these goals, while retaining a direct
connection to the microscopic Hamiltonian. Moreover, the
structure of the theory suggests that when the wavelengths of induced
inhomogeneties approach the microscopic scale, the response may be
challenging to understand in terms of dielectric response alone. This
suggests that one ought to exercise extreme caution when applying ideas from
continuum mechanics at the nanoscale.

\begin{acknowledgments}
S.J.C. is grateful to Michiel Sprik for sharing his insights on
electromechanics over several years. A.T.B. acknowledges funding from
the Oppenheimer Fund and Peterhouse College, University of Cambridge.
S.J.C. is a Royal Society University Research Fellow (Grant
No. URF\textbackslash R1\textbackslash 211144) at Durham University.
\end{acknowledgments}

\appendix

\section{The connection between hyper-DFT and LMFT applied to non-electrostatic interactions}
\label{sec:appendix-nonelectro}

We briefly discuss the connection between hyper-DFT and LMFT applied
to non-electrostatic interactions. We show that the result is
consistent with Archer and Evans \cite{Archer2013}, with additional
insights regarding the conditions placed on $\chi_{\rho,\rm
R}^{(\rho)}$.

We assume a pair interaction potential 
that can split into short-ranged and long-ranged parts
(using analogous notation to the main article):
\begin{equation}
w(r) = w_0(r) + w_1(r).
\end{equation}
A classical example in the theory of simple liquids is the
Weeks--Chandler--Anderson separation of the Lennard--Jones
potential \cite{Weeks1971}.
We also assert that the external potential operator
comprises two contributions,
\begin{equation}
\mcl{V}_{\rm ext}(\mbf{R}^N) = \sum^N_i V_{\rm ext}(\mbf{r}_i) + \sum^N_i \phi(\mbf{r}_i).
\end{equation}
Note that, in this appendix, $\phi$ has the dimensions of energy rather
than electrostatic potential. We now define the modified interparticle potential
energy as
\begin{equation}
    \hat{\mcl{U}}_{\rho}(\mbf{R}^N) =  \hat{\mcl{U}}(\mbf{R}^N) + \int\!\!\mrm{d}\mbf{r} \, \phi(\mbf{r})\hat{\rho}(\mbf{r},\{\mbf{r}_i\}).
\end{equation}
The derivation then follows that presented in Sec.~\ref{sec:LMFT} with
the substitutions: $n\to\rho$, $v_0\to w_0$, $v_1\to w_1$, and an
appropriate relabelling of the response functions and hyperdirect
correlation functionals. Note that in the hyper-DFT formulation of
LMFT, the two LMF conditions,
\begin{subequations}
\begin{align}
 c^{(1)}(\mbf{r};[\rho,\beta\phi]) - c_{\rm R}^{(1)}(\mbf{r};[\rho_{\rm R},\beta\phi_{\rm R}]) &=
 -\beta\Delta\mu, \\
 \rho^{(1)}(\mbf{r};[\rho,\beta\phi]) = \rho^{(1)}_{\rm R}(\mbf{r};[\rho,\beta\phi_{\rm R}]),
\end{align}
\end{subequations}
are distinct.

We find the following exact expressions relating $\Delta\phi$ to the first functional
derivatives of $\Delta\mcl{F}^{(\rm ex)}_{\rm intr}$,
\begin{widetext}
\begin{subequations}
\label{eqn:Fex-expand-final-appendix}
\begin{align}
\left.\frac{\delta\beta\Delta\mcl{F}^{(\rm ex)}_{\rm intr}[\varrho,\beta\phi]}{\delta\varrho(\mbf{r})}\right\vert_{\varrho=\rho}  &=
\beta\Delta\mu -\beta\Delta\phi(\mbf{r})
+ \mathcal{O}\big(\Delta\phi^2\big), \label{eqn:Fex-expand-final-rho-appendix} \\
\left.\frac{\delta\beta\Delta\mcl{F}^{(\rm ex)}_{\rm intr}[\rho,\beta\varphi]}{\delta\beta\varphi(\mbf{r})}\right\vert_{\varphi=\phi} &= 
\beta\int\!\mrm{d}\mbf{r}^\prime\,\chi_{\rho,\rm R}^{(\rho)}(\mbf{r},\mbf{r}^\prime;[\rho,\beta\phi])\Delta\phi(\mbf{r}^\prime)
+ \mathcal{O}\big(\Delta\phi^2\big), \label{eqn:Fex-expand-final-phi-appendix}
\end{align}
\end{subequations}
\end{widetext}
where we have noted that the hyperdirect correlation functions are now simply
\begin{equation}
c_{\rho,\rm R}^{\rm (\varrho\varphi)}(\mbf{r},\mbf{r}^\prime;[\rho,\beta\phi]) = c_{\rho,\rm R}^{\rm (\varphi\varrho)}(\mbf{r}^\prime,\mbf{r};[\rho,\beta\phi]) = \delta(\mbf{r}-\mbf{r}^\prime).
\end{equation}
\newpage
Through asserting an analogous mean field expression for
$\Delta\mcl{F}^{(\rm ex)}_{\rm intr}$ as Eq.~\ref{eqn:DeltaFex-MF}, i.e.,
\begin{widetext}
\begin{equation}
\label{eqn:DeltaFex-MF-appendix}
 \Delta\mcl{F}^{\rm (ex)}_{\rm intr}[\varrho,\beta\varphi] =
 \Delta\mu\int\!\mrm{d}\mbf{r}^\prime\,\varrho(\mbf{r}^\prime) +
  \frac{1}{2}\int\!\mrm{d}\mbf{r}^\prime\!\int\!\mrm{d}\mbf{r}^{\prime\prime}\,
 \delta_{\mrm{u}}\rho^{(1)}_{\rm R}(\mbf{r}^\prime;[\varrho,\beta\varphi])\delta_{\mrm{u}}\rho^{(1)}_{\rm R}(\mbf{r}^{\prime\prime};[\varrho,\beta\varphi])v_1(|\mbf{r}^\prime-\mbf{r}^{\prime\prime}|),
\end{equation}
\end{widetext}
where $\delta_{\mrm{u}}\rho^{(1)}_{\rm R}(\mbf{r};[\varrho,\beta\varphi])=\rho^{(1)}_{\rm R}(\mbf{r};[\varrho,\beta\varphi])-\rho_{\mrm{u,R}}$, 
with $\rho_{\mrm{u,R}}$ denoting the uniform bulk density of the reference fluid corresponding to $\mu_{\mrm{R}}$, we can write
\begin{widetext}
\begin{subequations}
\label{eqn:Fex-MF-final-appendix}
\begin{align}
\left.\frac{\delta\beta\Delta\mcl{F}^{(\rm ex)}_{\rm intr}[\varrho,\beta\phi]}{\delta\varrho(\mbf{r})}\right\vert_{\varrho=\rho} &=
\beta\Delta\mu +
\beta\int\!\mrm{d}\mbf{r}^{\prime\prime}\,\delta_{\mrm{u}}\rho^{(1)}_{\rm R}(\mbf{r}^{\prime\prime};[\rho,\beta\phi])w_1(|\mbf{r}-\mbf{r}^{\prime\prime}|), \label{eqn:Fex-MF-final-rho-appendix} \\
\left.\frac{\delta\beta\Delta\mcl{F}^{(\rm ex)}_{\rm intr}[\rho,\beta\varphi]}{\delta\beta\varphi(\mbf{r})}\right\vert_{\varphi=\phi} &=
-\beta\int\!\mrm{d}\mbf{r}^\prime\!\int\!\mrm{d}\mbf{r}^{\prime\prime}\,\chi_{\rho, \rm R}^{(\rho)}(\mbf{r},\mbf{r}^\prime;[\rho,\beta\phi])
\delta_{\mrm{u}}\rho^{(1)}_{\rm R}(\mbf{r}^{\prime\prime};[\rho,\beta\phi])w_1(|\mbf{r}^\prime-\mbf{r}^{\prime\prime}|).
\label{eqn:Fex-MF-final-phi-appendix}
\end{align}
\end{subequations}
\end{widetext}
Ignoring terms $\mathcal{O}(\Delta\phi^2)$ and higher in
Eqs.~\ref{eqn:Fex-expand-final-appendix}, and comparing to
Eqs.~\ref{eqn:Fex-MF-final-appendix} we find:
\begin{equation}
\label{eqn:DeltaPhi-appendix}
\Delta\phi(\mbf{r}^\prime) = -\int\!\!\mrm{d}\mbf{r}^{\prime\prime}\,\delta_{\mrm{u}}\rho^{(1)}_{\mrm{R}}(\mbf{r}^{\prime\prime};[\rho,\beta\phi]) w_1(|\mbf{r}^{\prime}-\mbf{r}^{\prime\prime}|).
\end{equation}
Eq.~\ref{eqn:Fex-MF-final-rho-appendix} is the same as Archer and
Evans, expressed in terms of a hyperdensity
functional. The additional insight we gain is in 
Eq.~\ref{eqn:Fex-MF-final-phi-appendix}, showing that the reference
system needs to be constructed such that $\chi_{\rho, \rm
R}^{(\rho)}(\mbf{r},\mbf{r}^\prime;[\rho,\beta\phi])$
ensures that the higher order terms can be neglected.

\section{Expression for $\Delta\mu$ for a polar fluid} \label{sec:polarmu}

As discussed in the main text, if $\kappa^{-1}$ is sufficiently large,
then the difference in bulk free energies between the true and mimic
systems will be dominated by differences in potential energy rather
than entropy. In Ref.~\onlinecite{Rodgers2009}, an analytical
correction for the Coulombic energy between the true and reference
bulk systems was derived, based on the Stillinger--Lovett moment
conditions \cite{Stillinger1968b}. Specifically, for a neutral polar fluid
\begin{equation}
    \Delta U = \langle\Delta\hat{\mcl{U}}\rangle = 
     \frac{N}{2\beta\rho_{\rm b}\kappa^{-3}\sqrt{\pi}^3}\frac{\epsilon - 1}{\epsilon} - \frac{2N p^2}{3\kappa^{-3}\sqrt{\pi}},
\end{equation}
where $\rho_{\mrm{b}} = N/V_{\rm b}$ is the average bulk density of
the system of volume $V_{\rm b}$, $p$ is the molecular dipole moment,
and $\epsilon$ is the dielectric constant of the fluid.  The change in
internal energy of the bulk system is
\begin{equation}
    \mrm{d}U = T\mrm{d}S + \mu\mrm{d}N  - P\mrm{d}V,
\end{equation}
where $S$ is the entropy, $N$ is the number of molecules and $P$ is the pressure.
The chemical potential is then
\begin{equation}
        \mu = \left( \frac{\partial U}{\partial N} \right)_{V,T} - T\left( \frac{\partial S}{\partial N} \right)_{V,T}.
\end{equation}
Assuming that $\kappa^{-1}$ is large enough that $S\approx
 S_{\mrm{R}}$, the correction for the chemical potentials is
\begin{equation}
\label{eqn:DeltaMu-dielectric}
\begin{split}
 \Delta \mu  &\equiv \mu - \mu_{\mrm{R}} = \left( \frac{\partial \Delta U}{\partial N} \right)_{V,T} \\
 &= \frac{1}{2\beta\rho_{\rm b}\kappa^{-3}\sqrt{\pi}^3}\frac{\epsilon - 1}{\epsilon} - \frac{2 p^2}{3\kappa^{-3}\sqrt{\pi}}.
\end{split}
\end{equation}
Note that, for an ionic fluid, 
the chemical potential shifts are not given 
by the $\epsilon \rightarrow \infty$ limit of Eq.~\ref{eqn:DeltaMu-dielectric}. In this case,
for ionic species $\gamma$ with charge $q_\gamma$
\begin{equation}
    \Delta \mu_{\gamma} = \frac{-q^2_{\gamma}}{\kappa^{-1}\sqrt{\pi}},
\end{equation}
as derived in Ref.~\cite{bui2024learningclassicaldensityfunctionals}.

\section{Discussion on how $\chi_{n,\rm R}^{(n)}$ is related to the dielectric constant} \label{sec:dielectricconstant}

An appealing feature of density functional approaches is their
thermodynamic consistency. As the dielectric constant appears in
Eq.~\ref{eqn:DeltaMu-dielectric}, it would therefore be appealing if
the same dielectric constant applies to the short-ranged reference
system. As $\epsilon$ is an intensive material property that is
determined by short-ranged correlations
\cite{Kirkwood1939,Madden1984}, such an assertion would in
fact be reasonable. Further evidence to support this notion can be
found in Ref.~\cite{Cox2020}, in which the longitudinal 
fluctuations of
the polarization measured by
$4\pi\chi^{(0)}_{\rm R, zz}$, with $\chi^{(0)}_{\rm R, zz}$
denoting the longitudinal susceptibility, were 
found to tend to $\epsilon-1$ in the
short-ranged reference system as opposed to the expected
$(\epsilon-1)/\epsilon$. 
Here we
derive an analytic expression for the Fourier components of $\chi_{n,\rm
R}^{(n)}$ in terms of $\epsilon$, $\kappa$, and $\beta$. For simplicity,
we will present the derivation for a bulk isotropic fluid 
and consider $\chi_{n,\rm R}^{(n)}$ as a response function, dropping
its formal functional dependence.

We begin with the established result \cite{HansenMcDonaldBook}
for the true system,
\begin{equation}
 \lim_{k\to 0}\frac{4\pi\beta}{k^2}\tilde{\chi}^{(n)}_n(k) = \frac{\epsilon-1}{\epsilon}
\label{eqn:dielectricfunction}
\end{equation}
where the tilde indicates the Fourier component of a function. Taking
the Fourier transform of Eq.~\ref{eqn:phiR-weeks} gives,
\begin{equation}
   \beta \tilde{\phi}_{\mrm{R}}(\mbf{k}) =  \beta\tilde{\phi}(\mbf{k}) + \frac{4\pi\beta}{k^2}\exp\left(-\frac{k^2}{4\kappa^2}\right)\tilde{n}_{\rm R}(\mbf{k}).
\end{equation}
The charge densities and the potentials are related via the
``charge--charge'' response functions
\begin{subequations}
\begin{align}
  \tilde{n}(\mbf{k})   &= -\beta\tilde{\chi}_{n}^{(n)}(\mbf{k})\,\tilde{\phi}(\mbf{k}), \\
  \tilde{n}_{\rm R}(\mbf{k}) &= -\beta\tilde{\chi}_{n,\rm R}^{(n)}(\mbf{k})\,\tilde{\phi}_{\rm R}(\mbf{k}).
\end{align}
\end{subequations}
Enforcing the LMF condition for the charge densities
in Eq.~\ref{eqn:LMF-2}, we arrive at the
following expression,
\begin{equation}
  \frac{4\pi\beta}{k^2}\tilde{\chi}^{(n)}_{n, \mrm{R}}(k) = \dfrac{\frac{4\pi\beta}{k^2}\tilde{\chi}^{(n)}_n(k)}{1-\frac{4\pi\beta}{k^2}\tilde{\chi}^{(n)}_n(k)\exp(-\frac{k^2}{4\kappa^2})}.
\end{equation}
Note that for sufficiently large $k$,
$\hat{\chi}^{(n)}_{n, \mrm{R}}(k) \approx \hat{\chi}^{(n)}_{n}(k)$. However,
agreement occurs at much longer wavelengths that the naive estimate
$k \approx 2\pi\kappa$.

Symmetry arguments dicate that \[\lim_{k\to 0}
4\pi\beta\tilde{\chi}^{(n)}_{n}(k)/k^2 = \frac{\epsilon-1}{\epsilon}
+ \mathcal{O}(k^2).\]

Assuming that $\kappa^{-1}$ is sufficiently large, the behavior of
$4\pi\beta\tilde{\chi}^{(n)}_{n,\rm R}(k)/k^2$ as $k\to 0$ will then
be dominated by the exponential,
\begin{equation}
    \lim_{k\rightarrow 0 } \frac{4\pi\beta}{k^2}\tilde{\chi}^{(n)}_{n, \mrm{R}}(k)
    = \dfrac{(\epsilon-1)/\epsilon}{1-\left(\frac{\epsilon-1}{\epsilon}\right)\exp(-\frac{k^2}{4\kappa^2})}=\epsilon-1.
    \label{eqn:chinn_R}
\end{equation}
This result is consistent with the notion that $\epsilon$ is
determined by short-ranged correlations; in the short-ranged system,
longitudinal polarization fluctuations tend toward the same value as
their unscreened transverse counterparts.

\bibliography{references}

\end{document}